\documentclass[%
 reprint,
superscriptaddress,
nofootinbib,
 amsmath,amssymb,
 aps,
]{revtex4-1}
\usepackage{graphicx}
\usepackage{dcolumn}
\usepackage{bm}
\usepackage{color}
\usepackage[colorlinks=true,linkcolor=blue,citecolor=blue]{hyperref}
\usepackage[mathlines]{lineno}
\usepackage{csquotes}
\usepackage{lipsum}

\begin{document}

\preprint{APS/123-QED}

\title{Ab initio multi-shell valence-space Hamiltonians and the island of inversion}

\author{T.~Miyagi}
 \affiliation{TRIUMF, 4004 Wesbrook Mall, Vancouver, BC V6T 2A3, Canada}
\author{S.~R.~Stroberg}%
\affiliation{Department of Physics, University of Washington, Seattle, WA 98195, USA}

\author{J.~D.~Holt}
\affiliation{TRIUMF, 4004 Wesbrook Mall, Vancouver, BC V6T 2A3, Canada}%
\affiliation{Department of Physics, McGill University, 3600 Rue University, Montr\'eal, QC H3A 2T8, Canada}

\author{N.~Shimizu}
\affiliation{Center for Nuclear Study, The University of Tokyo, 7-3-1, Hongo, Bunkyo-ku, Tokyo, 113-0033, Japan  }%

\date{\today}

\begin{abstract}
In the shell-model framework, valence-space Hamiltonians connecting multiple major-oscillator shells are of key interest for investigating the physics of neutron-rich nuclei, which have been the subject of intense experimental activity for decades.
Here we present an extension of the ab initio valence-space in-medium similarity renormalization group which allows the derivation of such Hamiltonians nonperturbatively.
Starting from initial two- and three-nucleon forces from chiral effective field theory, we then calculate properties of nuclei in the important island-of-inversion region above oxygen, so far unexplored with ab initio methods.
Our results in the neon and magnesium isotopes indicate the importance of neutron excitation from the $sd$ to $pf$ shells and ground states dominated by intruder configurations around $N=20$, consistent with the conclusions from phenomenological studies.
We also benchmark the excitation spectrum of $^{16}$O with coupled-cluster theory, finding generally good agreement, and discuss implications for ground state energies and charge radii in oxygen and calcium isotopes.
Finally we outline the proper procedure for treating the long-standing issue of center-of-mass contamination, and show that with a particular choice of valence space, these spurious states can be removed successfully.
\end{abstract}

\maketitle


\section{\label{sec:introduction}Introduction}
Recent efforts in low-energy nuclear structure have been particularly focused on unstable exotic nuclei, the study of which is the primary physics motivation driving the development of next-generation rare-isotope-beam facilities around the world.
These nuclei are important for understanding long-standing problems, such as astrophysical nucleosynthesis~\cite{Mumpower2016,Martin2016}, the formation and evolution of nuclear magic numbers~\cite{wienholtz2013,Steppenbeck2013,Taniuchi2019}, and the limits of the nuclear landscape~\cite{Sakurai1999,Otsuka2010,Erler2012,Holt2019,Ahn2019,Neufcourt2020}, with an eye towards a comprehensive understanding of all nuclei.
One key phenomenon arising in these studies is the \enquote{island of inversion}~\cite{Warburton1990,Caurier2014}, where standard magic numbers break down, and nuclei in these regions no longer exhibit expected ``magic'' properties.
Tremendous effort has been made both in experiment~\cite{Thibault1975,
Detraz1979,Guillemaud-Mueller1984,
Motobayashi1995,Iwasaki2001,
Yordanov2007,Wimmer2010,Doornenbal2013,Doornenbal2016} and and theory~\cite{
Campi1975,Poves1987,Caurier1998,Utsuno1999,
Yamagami2004,Tsunoda2017} to understand, in particular,
the disappearance of the $N=20$ neutron magic number in the neon through magnesium ($Z$=10--12) isotopes.
While this phenomenon is typically interpreted in terms of enhanced particle-hole excitations from $sd$ to $pf$ orbits, driven by the decreased shell gap between the major oscillator shells, so far all such studies have been based on phenomenological or quasi-microscopic models.

A fundamental goal of ab initio nuclear theory is to understand the properties and structure of nuclei from the underlying interactions between nucleons, with no input from data beyond that necessary to inform theories of nuclear forces.
The valence-space formulation of the in-medium similarity renormalization group (VS-IMSRG) is a broadly applicable ab initio many-body method capable of providing an array of observables of ground and excited states of essentially all open-shell nuclei accessible to the traditional shell model, including energies, charge radii, and electroweak moments and transitions~\cite{Tsukiyama2012,Bogner2014, Stroberg2016, Hergert2016, Stroberg2017, Parzuchowski2017a, Stroberg2019,Holt2019,Gysb19GT}.
In the standard phenomenological shell model, one- and two-body matrix elements of this Hamiltonian are fit to reproduce experimental data.
In the VS-IMSRG, as detailed below, we instead start from two- (NN) and three-nucleon (3N) forces to non-perturbatively decouple an effective valence-space Hamiltonian.
In the absence of approximations, diagonalization of this Hamiltonian would reproduce eigenvalues of the full $A$-body problem.

Despite these successes, the derivation of effective Hamiltonians for valence spaces spanning multiple major shells, essential for studying the island of inversion region, is challenging due to the problem of intruder states~\cite{Schucan1972,Schucan1973,Stroberg2019}.
One recent approach based on many-body perturbation theory, the extended Kuo-Krenciglowa method, has shown promise for generating multi-shell Hamiltonians perturbatively~\cite{Tsunoda2014a,Tsunoda2017}, but single-particle energies are still fit to provide a good description of data in the region.
A complementary approach currently being pursued is to target a specific state within a multi-reference framework~\cite{Yao2018}.

In this paper we extend the reach of the VS-IMSRG to decouple valence space Hamiltonians across multi-major shells to provide the first ab initio description of the island of inversion in the region above oxygen.
Through the introduction of a new generator of the transformation, we show that previously intractable cases now decouple smoothly.
Nevertheless, in practice, we find that center-of-mass contamination remains problematic.
We therefore discuss a procedure for removing these spurious states in ab initio valence-space approaches and show that this can be achieved for particular choices of multi-shell spaces.
We use these Hamiltonians to explore the physics of the island of inversion in the neon, magnesium, and silicon isotopes, showing that the expected interplay between normal and intruder orbits in ground and excited states can be successfully described within our approach.
We then turn to other interesting applications such as the computation of the excitation spectrum of doubly magic $^{16}$O, benchmarking against coupled-cluster theory~\cite{Ekstrom2015} and experiment, finding overall good agreement.
Finally, we investigate the trends of ground state energies and charge radii in the oxygen and radii isotope shift in calcium isotopic chains.

\section{\label{sec:imsrg}IMSRG and novel multishell generator}
In the valence-space framework, the single-particle Hilbert space is partitioned into core, valence, and \enquote{outside} spaces.
In the final calculation, the core (e.g., $^{4}$He, $^{16}$O, $^{40}$Ca) and outside orbits are taken to be inactive, and a Hamiltonian in the valence space, containing the essential degrees of freedom to reproduce the low-lying states, is where we consider all possible configurations through exact diagonalization.
Our goal here is to construct an effective Hamiltonian where excitations out of the core or into the outside space are explicitly decoupled.
To this end, in the IMSRG we evolve the Hamiltonian using the flow equation:
\begin{equation}\label{eq:IMSRGflow}
    \frac{dH(s)}{ds} = [\eta(s), H(s)],
\end{equation}
where the Hamiltonian, which depends on the flow parameter $s$, is expressed in second-quantized form:
\begin{align}
     H(s) &= E_{0}(s) + \sum_{ab} f_{ab}(s) \{a^{\dag}_{a} a_{b}\}
     \notag \\ & \hspace{1em}
 + \frac{1}{4} \sum_{abcd} \Gamma_{abcd} \{a^{\dag}_{a}a^{\dag}_{b}a_{d}a_{c}\}.
\end{align}
$E_{0}$, $f_{ab}$, and $\Gamma_{abcd}$ are zero-, one-, two-body matrix element of Hamiltonian, respectively.
The operator $a_{a}$ ($a^{\dag}_{a}$) annihilates (creates) the particle in the orbit $a$,
and $\{\ldots\}$ indicates normal ordering with respect to the single-determinant or ensemble reference state~\cite{Stroberg2017}.
The anti-Hermitian operator $\eta(s)$ is known as the generator:
\begin{equation}
\label{eq:eta}
    \eta = \sum_{ai}\eta_{ai}\{a^{\dagger}_{a}a_i\} + \sum_{abij}\eta_{abij}\{a^{\dagger}_aa^{\dagger}_ba_ja_i\} - {\rm h.c.}
\end{equation}
with
\begin{equation}
    ai\in \{pc,~ov\},\hspace{1.5em} abij\in\{pp'cc',~pp'vc,~opvv'\}.
\end{equation}
The indices $c$, $v$, and $o$ indicate core, valence, and outside-space orbits, respectively, and $p$ indicates either $v$ or $o$.

In this work, for purposes of deriving effective Hamiltonians across multiple major shells, we define a new generator:
\begin{align}
\eta_{ai}&= \frac{1}{2} \arctan \left(
\frac{2f_{ai}}{f_{aa} - f_{ii} + \Gamma_{aiai} + \Delta}\right),\label{eq:eta1b} \\
\eta_{abij} &= \frac{1}{2} \arctan\left(
\frac{2\Gamma_{abij}}{f_{aa} + f_{bb} - f_{ii} - f_{jj} + G_{abij} + \Delta}
\right), \label{eq:eta2b}\\
G_{abij} &= \Gamma_{abab}+\Gamma_{ijij}-(\Gamma_{aiai}+\Gamma_{bjbj}+[a\leftrightarrow b]).
\end{align}
Here we have introduced the energy denominator shift $\Delta$, which solves issues inherent in decoupling multi-shell Hamiltonians, as discussed in more detail below.
Note that our choice is the same as the generator used in many earlier
works~\cite{Bogner2014,Stroberg2016, Stroberg2017} except for the energy shift $\Delta$.
Adding $\Delta$ can be regarded as simply taking another generator, similar to choosing from the standard Wegner, White, or Imaginary-time generators used in IMSRG calculations~\cite{Hergert2016}.
Instead of directly integrating the flow equation \eqref{eq:IMSRGflow}, we use the Magnus formulation of the IMSRG~\cite{Morris2015}.
This approach explicitly produces the unitary transformation, which enables a much more efficient treatment of observables~\cite{Parzuchowski2017a}.
Finally, the IMSRG evolution induces three- and higher-body terms which should be kept in principle, but are inconvenient in practice.
Here, we keep up to two-body terms (known as IMSRG(2) approximation), which has been observed to be an effective many-body truncation in many cases~\cite{Tsukiyama2011,Tsukiyama2012,Hergert2013,Hergert2013a,Hergert2014,Hergert2016,Hergert2017,Morris2018}.

\section{\label{sec:results} Numerical Analyses}
Throughout this work, our calculations are done with NN and 3N interactions derived from chiral effective field theory~\cite{Epelbaum2009,Machleidt2011}.
We work in the harmonic oscillator basis, with frequency $\hbar\omega=16$ MeV, defined by $e_{\max} = \max(2n+l)$, where $n$ and $l$ are the radial quantum number and angular momentum,
respectively.
The full treatment of three-body matrix element is challenging due the memory limitations, so we apply the additional truncation
$ E_{3\max} = \max(2n_{1}+l_{1}+2n_{2}+l_{2}+2n_{3}+l_{3})$.
The IMSRG calculations to generate the effective valence-space Hamiltonians, radii, and $E2$ operators were performed with the imsrg++ code~\cite{Stroberg}, and the final diagonalization within the valence space and calculations for corresponding transition densities are done with the KSHELL code~\cite{Shimizu2019}.

\subsection{\label{sec:delta} Effects of the energy denominator shift}

\begin{figure}[t]
    \centering
    \includegraphics[clip,width=\columnwidth]{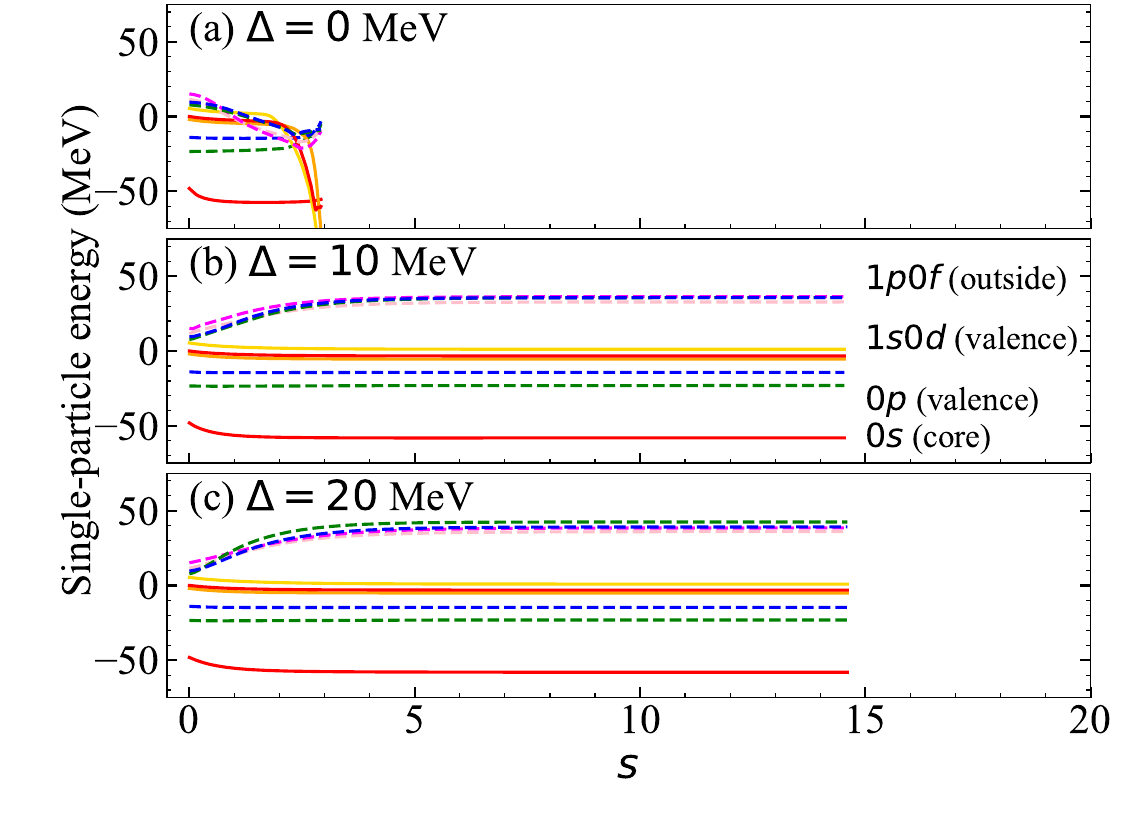}
    \caption{Flow of the single-particle energies for $^{16}$O ($^{4}$He core,
    $psd$ valence space) for
    $\Delta=0$ (a), $10$ (b), and $20$ MeV (c) cases.
    The solid (dashed) curves show the flow of positive (negative)
    parity levels. All the calculations are done in the $e_{\max}=8$ model space.}
    \label{fig:flow_spe}
\end{figure}

To illustrate the role of the energy denominator shift $\Delta$, in Fig.~\ref{fig:flow_spe} we show the flow of the neutron single-particle energies, i.e. $f_{a}(s)\equiv f_{aa}(s)$, for $0s$ (core), $0p$ and $1s0d$ (valence), and $1p0f$ (outside) orbits for a calculation decoupling a $psd$ valence space, using an $^{16}$O reference.
With $\Delta=0$ MeV [Fig.~\ref{fig:flow_spe} (a)], we find that as the flow parameter $s$ increases, the trajectory of some outside levels causes them to drop below the valence-space levels, and flow does not converge.
Note that $\Delta=0$ MeV corresponds to the generator used in the earlier  works~\cite{Stroberg2016,Stroberg2017}, and similar unstable patterns were occasionally observed even in cases of single-shell valence-space decoupling.
According to the figure, the ill-behaved flow begins where $d f_{o}/ds < 0$, where, $f_{o}$ is a single-particle energy of an outside orbit.
We can understand how this quantity can become negative by inspecting the one-body part\footnote{To simplify the argument, we assume the Hartree-Fock basis, so $f_{ab}$ is diagonal, and take $\eta$ to be the White generator with Moller-Plesset energy denominators (replace $\arctan(x)\to x$ and remove $\Gamma$ and $G$ from the denominators in \eqref{eq:eta1b} and \eqref{eq:eta2b}).} of the flow equation~\eqref{eq:IMSRGflow} ~\cite{Hergert2016}:
\begin{align}\label{eq:foflow}
     \frac{d f_{o}}{ds} &= \sum_{p}\sum_{hh'}
     \frac{|\Gamma_{ophh'}|^{2}}{f_{o} + f_{p} - f_{h} - f_{h'}} \notag \\
    &\hspace{2em}+ \sum_{pp'}\sum_{h} \frac{|\Gamma_{ohpp'}|^{2}}{f_{o} + f_{h} - f_{p} - f_{p'}}.
\end{align}
Here, the subscripts $h$ and $p$ indicate the occupied and unoccupied orbits, respectively.
Since the numerators of both terms in~\eqref{eq:foflow} are manifestly positive, the left hand side can only be negative when at least one of the energy denominators are negative.
This could happen for the second term if, e.g. $f_{0f_{7/2}}+f_{0p_{3/2}}-2f_{0d_{3/2}}$ were negative.
The denominator and $df_o/ds$ become more negative with increasing $s$ and the flow diverges.
As a rule of thumb, we can expect problems in the flow whenever the range of single-particle energies within a valence space is larger than the gaps between valence and core/outside orbits.
Similar criterion involving the intruder states were discussed through the convergence of the perturbation series~\cite{Schucan1972}.

To avoid the negative energy denominator we can add an energy shift $\Delta$, which maintains the decoupled fixed-point of the SRG flow.
The size of $\Delta$ should be comparable to the single-particle-energy gap, which naively can be empirically estimated as $\sim$ 41$A^{-1/3}$ MeV.
Then, the suitable value of $\Delta$ should be on the order of 10 MeV.
In Figs.~\ref{fig:flow_spe} (b) and (c), we show $\Delta=10$ and $20$ MeV cases.
 As seen in these figures, the \emph{outside} levels rise with increasing $s$, and the flow
 is stable.
We have checked that the dependence of computed energies on the value of $\Delta$ is small (several tens keV change between $\Delta=$ 10 and 20 MeV).
Therefore, following discussions are based on the calculations at $\Delta=10$ MeV.

\subsection{\label{sec:com} Center-of-mass motion}

In multi-shell valence-space Hamiltonians, whether phenomenologically or microscopically constructed, the issue of center-of-mass (c.m.) contamination must be treated carefully.
While the Hamiltonian depends solely on intrinsic coordinates, we work in the laboratory frame, which combines intrinsic and c.m. coordinates.
The resulting wave function $|\Psi\rangle$ is in general \mbox{$|\Psi\rangle = \sum_{iI} c_{iI} |\Psi^{\rm intr}_{i}\rangle \otimes |\Psi^{\rm c.m.}_{I}\rangle$}, with the intrinsic wave function $|\Psi^{\rm intr}_{i} \rangle$, c.m. wave function $|\Psi^{\rm c.m.}_{I}\rangle$, and coefficient $c_{iI}$.
As discussed in the context of coupled-cluster calculations~\cite{Hagen2009a}, we expect that for a sufficiently large laboratory-frame model space, only one $c_{iI}$ is dominant, i.e. $|\Psi\rangle \approx |\Psi^{\rm intr}_{i} \rangle \otimes |\Psi^{\rm c.m.}_{I} \rangle$.
Assuming this factorization, we can use the Gl\"ockner-Lawson prescription~\cite{Gloeckner1974} to remove spurious excited states due to the c.m. motion.

This approach is often used in conventional shell-model calculations, and while stability with respect to the Gl\"ockner-Lawson term is typically observed, this prescription is not strictly valid for two reasons.
First, the single-particle energies of the conventional valence-space Hamiltonian are not consistent with the harmonic-oscillator energy spectra, i.e., the single-particle energies within a major shell are not degenerate.
Second, the Gl\"ockner-Lawson term indtroduces undesired coupling between the valence-space and outside orbitals.
There is an alternative option when we add the $H_{\rm c.m.}$ operator.
Since the Hamiltonian is no longer expressed in the harmonic oscillator basis after the VS-IMSRG transformation,  the $H_{\rm c.m.}$ should be consistently transformed.
Even if we transform $H_{\rm c.m.}$ consistently, adding transformed $H_{\rm c.m.}$ spoils the decoupling accomplished by the VS-IMSRG transformation obtained with solely $H_{\rm intr}$.
For these reasons, our adopted procedure for treating c.m. motion is to add the Gl\"ockner-Lawson term at the beginning of the calculation, where the harmonic oscillator basis representation is initially used.
In this case our Hamiltonian is $H = H_{\rm intr} + \beta H_{\rm c.m.}$, where $H_{\rm intr}$ is the intrinsic Hamiltonian, $H_{\rm c.m.} = {\mathbf P}^{2}/2Am + mA\tilde{\omega}^{2}{\mathbf R}^{2}/2 - 3\hbar\tilde{\omega}/2$ is the c.m. Hamiltonian, with the scaling parameter $\beta$ and $\tilde{\omega}$ the frequency of the c.m. Gaussian wave function.
Note that the expectation value $\langle H_{\rm c.m.} \rangle$ vanishes if the c.m. wave function can be factorized by a single Gaussian specified by $\tilde{\omega}$.
In general $\tilde{\omega}$ is not the same as the basis frequency $\omega$ and has to be optimized for each calculated state in the same manner discussed in Ref.~\cite{Hagen2010}.
As we discuss later, $\langle H_{\rm c.m.} \rangle$ approaches zero with increasing $\beta$, and therefore, we use $\tilde{\omega} = \omega$ in this work.

To summarize, our calculation procedure is the following.
We begin with $H = H_{\rm intr} + \beta H_{\rm c.m.}$ expressed in the harmonic oscillator basis.
Then, the single-particle orbits are optimized through the Hartree-Fock calculation with the ensemble normal ordering~\cite{Stroberg2017}.
By keeping up to two-body terms, the VS-IMSRG transformation is computed with the newly introduced generator [see Eqs.~(\ref{eq:eta})-(\ref{eq:eta2b})].
Finally, we solve the valence-space problem with the VS-IMSRG transformed operators.
In the practical calculations, we increase $\beta$ to push the c.m. modes up out of the spectrum until the calculated energies become $\beta$-independent.

\begin{figure}[t]
    \centering
    \includegraphics[clip,width=\columnwidth]{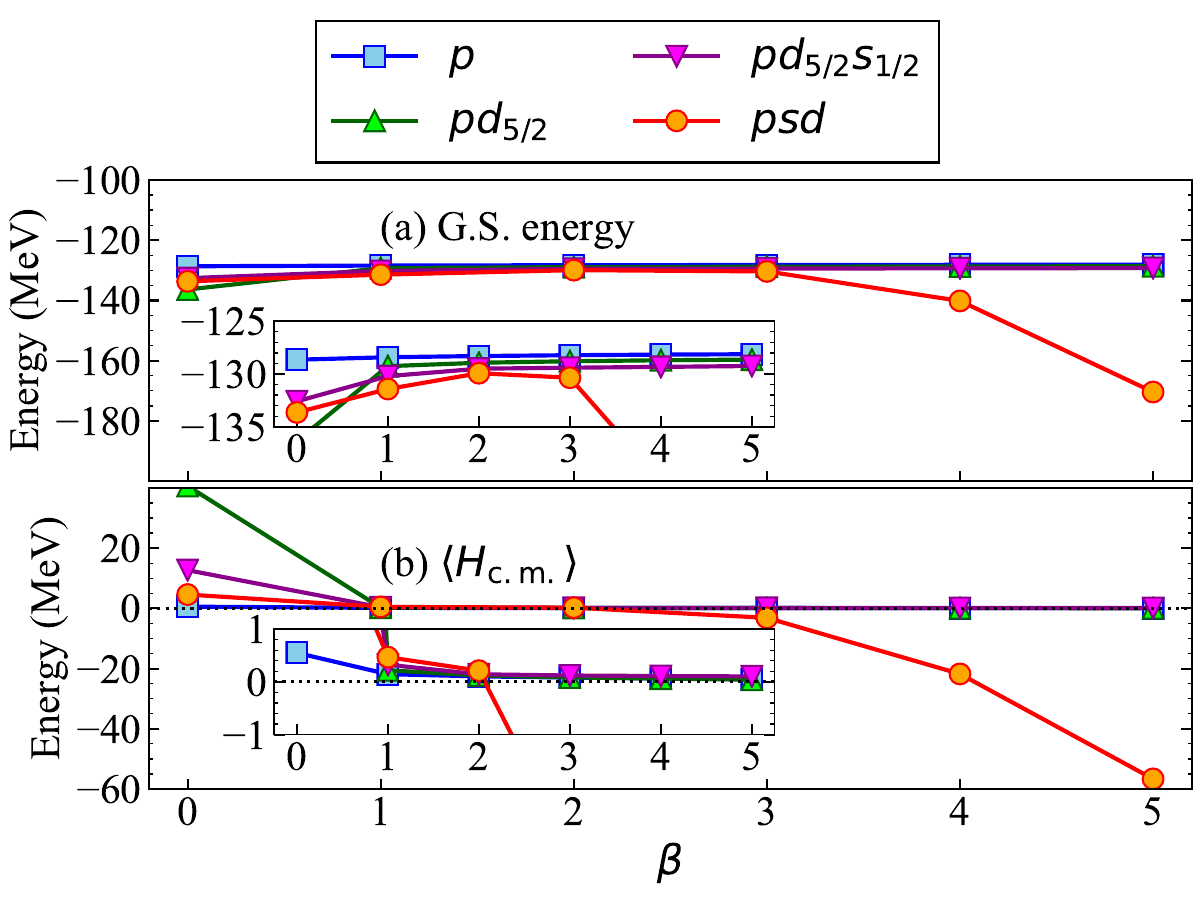}
    \caption{Ground-state energies (a) and expectation value of the
    c.m. Hamiltonian (b) for $^{16}$O calculated from the
    $^{4}$He-core $p$, $pd_{5/2}$, $pd_{5/2}s_{1/2}$,
    and $psd$ valence spaces.
    All the calculations are done in the $e_{\max}=8$ model space.
    The inset panels show an enlarged view.}
    \label{fig:O16_beta}
\end{figure}

In Fig.~\ref{fig:O16_beta} (a) we show the ground-state energy of $^{16}$O, taking $^{4}$He as the core and several choices of valence space: $p$, $pd_{5/2}$, $pd_{5/2}s_{1/2}$, or $psd$.
We see that the $p$, $pd_{5/2}$, and $pd_{5/2}s_{1/2}$ valence-space energies are nearly $\beta$-independent, while the $psd$ valence-space exhibits a strong $\beta$-dependence for $\beta \gtrsim 4$.
To examine further, we calculate the expectation values of the c.m. Hamiltonian $\langle H_{\rm c.m.} \rangle$ shown in Fig.~\ref{fig:O16_beta} (b).
For the $p$, $pd_{5/2}$, and $pd_{5/2}s_{1/2}$ valence-space calculations, $\langle H_{\rm c.m.} \rangle$ are almost zero, which is consistent with a single Gaussian for the c.m. wave function.
However, for the $psd$ valence-space, energies are shifted down as $\beta$ increases.
Since $H_{\rm c.m.}$ is originally positive definite, the large negative value of $\langle H_{\rm c.m.} \rangle$ implies that the induced many-body terms of $H_{\rm c.m.}$ are large.
Combined with the energy lowering in $\beta \gtrsim 4$, it is expected that the induced many-body terms of $H$ are not small, and the IMSRG(2) approximation breaks down.
Although adding the c.m. Hamiltonian is needed to remove the spurious modes, it breaks the hierarchy of induced terms in some cases.
As seen in Fig.~\ref{fig:O16_beta}, the choice of valence space is crucial to mitigate the effects of these induced many-body terms. In the following applications, we only show calculations where these induced terms are under control and illustrate the typically minor impact of changing $\beta$.

\section{\label{sec:applications} Results and Applications}

\subsection{\label{sec:NeMg} Island of inversion}

\begin{figure}[t]
    \centering
    \includegraphics[clip,width=\columnwidth]{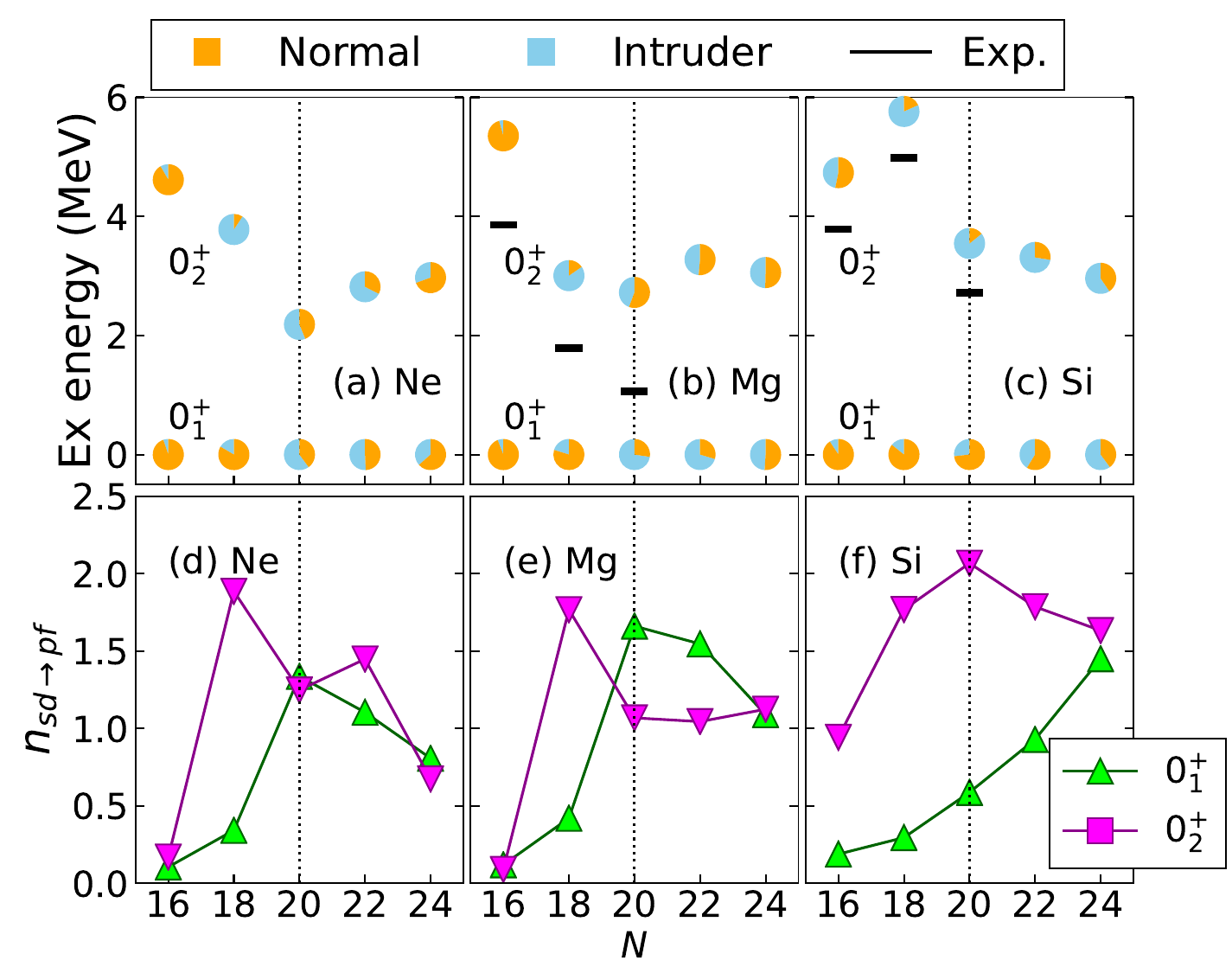}
    \caption{$0^{+}_{2}$ excitation energies and number of exciting neutrons from $sd$ to $pf$ orbits for neon [(a) and (d)], magnesium [(b) and (e)], and
    silicon [(c) and (f)].
    The valence-space Hamiltonian is derived for the $\pi sd$,$\nu sdf_{7/2}p_{3/2}$ space.}
    \label{fig:IoI}
\end{figure}

 \begin{figure*}[t]
    \centering
    \includegraphics[clip,width=\linewidth]{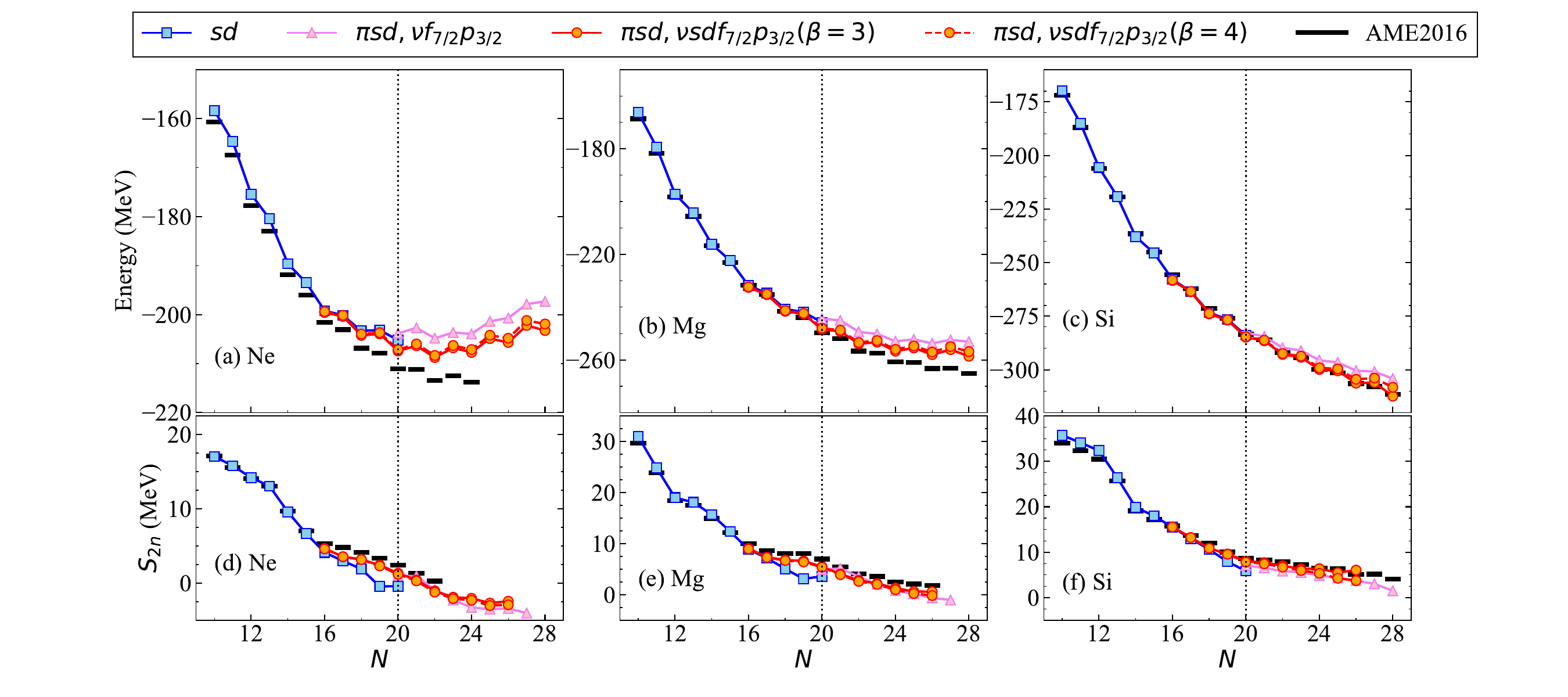}
    \caption{Ground-state energies [(a), (b), and (c)] and two-neutron separation energies  [(d), (e), and (f)] for neon, magnesium, and silicon isotopes.
    Dotted vertical lines indicate the border of $sd$- and $pf$-shell spaces.
    The experimental data (black bars) are taken from atomic mass evaluation 2016~\cite{Wang2017}.
    All the calculations are done with $e_{\max}=12$ and
    $E_{3\max}=16$.}
    \label{fig:nemgsi_gs}
\end{figure*}

\begin{figure*}[t]
    \centering
    \includegraphics[clip,width=\linewidth]{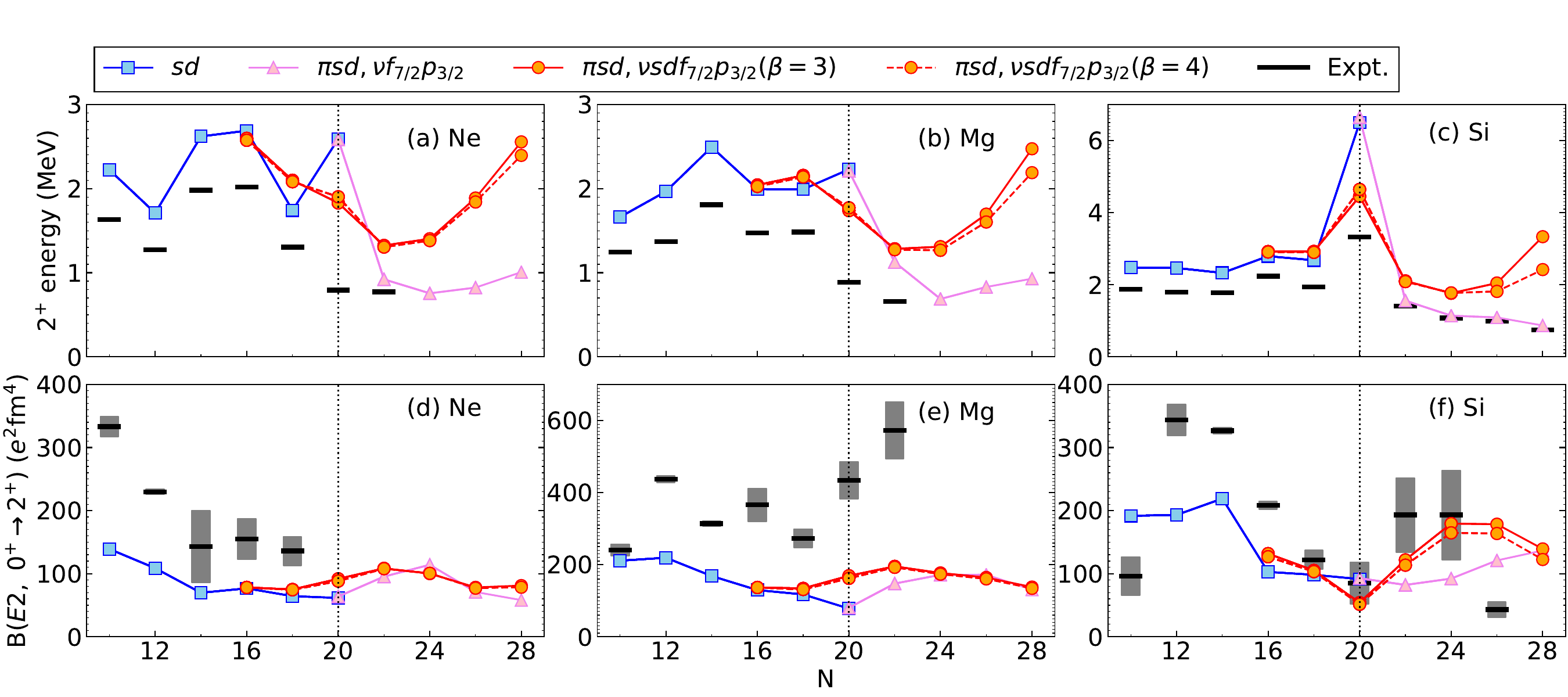}
    \caption{Excitation energies of and transition probability
    to first $2^{+}$ state for even-even neon [(a) and (d)]
    magnesium [(b) and (e)], and silicon isotopes [(c) and (f)].
    Dotted vertical lines indicate the border of $sd$- and $pf$-shell spaces.
    The experimental data (black bars) are taken from nuclear data center~\cite{NationalNuclearDataCenter}.
    All the calculations are done with $e_{\max}=12$ and
    $E_{3\max}=16$.}
    \label{fig:NeMgSi_Ex}
\end{figure*}

As the primary application of this development, we first discuss the well-known island of inversion in the neon ($Z$=10), magnesium ($Z$=12), and silicon ($Z$=14) isotopes.
We take for our valence space the neutron $0f_{7/2}$ and $1p_{3/2}$ orbits in addition to the standard $sd$ space above a $^{16}$O core, referred as $\pi sd, \nu sdf_{7/2}p_{3/2}$.
For our input nuclear NN+3N Hamiltonian, we use the 1.8/2.0(EM) interaction of Refs.~\cite{Hebeler2011,Simonis2016,Simonis2017}, which was obtained by combining a free-space SRG evolved $NN$ with an unevolved $3N$ interactions~\cite{Entem2003,Hebeler2011}.
This interaction well reproduces ground-state energies to A $\sim$ 100 region~\cite{Morris2018,Stroberg2019}, specifically with an rms deviation from experiment of roughly 3.5 MeV across the light and medium-mass regions~\cite{Holt2019}.
The IMSRG evolution is performed in a space with $e_{\max}=12$ and $E_{3\max}=16$.

In the top row of Fig.~\ref{fig:IoI}, we show the excitation energies of $0^{+}_{2}$ states in neon, magnesium, and silicon in the vicinity of $N=20$ as a function of neutron number.
Compared with the experimental excitation energies, we see that our energies are systematically high.
This is likely due to the IMSRG(2) approximation employed in this work, as similar trend have been observed in earlier studies~\cite{Simonis2017,Morris2018,Taniuchi2019}.
For each state we also illustrate the contribution of normal and intruder (defined by neutron holes in the $sd$ shell) configurations to the wave functions.
We see in Fig.~\ref{fig:IoI} (a) and (b), that the intruder component in the ground state is  strongly enhanced as we move towards $N = 20$ in neon and magnesium.
In contrast, the ground state in the silicon isotopes is dominated by normal configurations, as shown in Fig.~\ref{fig:IoI} (c).

In the lower row, we show the number of excited neutrons from the $sd$ to $pf$ shells ($n_{sd \rightarrow pf}$) for the ground ($0^{+}_{1}$) and $0^{+}_{2}$ states of the same isotopic chains.
Similar to an earlier study based on a phenomenological valence-space Hamiltonian~\cite{Utsuno1999}, we see a sudden jump in $n_{sd \rightarrow pf}$ for the ground states of neon and magnesium at $N=20$.
Furthermore, the inversion of the $n_{sd \rightarrow pf}$ is clearly observed between the ground and excited states, highlighting the breakdown of the $N=20$ shell gap in these isotopes.
On the other hand, the very modest increase of $n_{sd \rightarrow pf}$ in the ground state of the silicon isotopes indicates the persistence of the $N=20$ gap.
This overall behavior of $n_{sd \rightarrow pf}$ is in contrast to those obtained with a semi-microscopic many-body perturbation theory approach~\cite{Tsunoda2017}, which found a much more gradual transition into the island of inversion region.
We note, however, that the quantity $n_{sd \rightarrow pf}$ is not observable, and depends on the chosen Hamiltonian, so strong conclusions should not be drawn based on levels of agreement.

To see the impact of the neutron excitations from $sd$ to $pf$, we also employ the $^{16}$O-core $sd$ and $^{28}$O-core $\pi sd, \nu f_{7/2}p_{3/2}$ valence spaces.
In Fig.~\ref{fig:nemgsi_gs} panels (a), (b), and (c), we show the ground-state energies of neon, magnesium, and silicon isotopes, respectively, calculated within these different valence spaces.
As we move to the neutron rich region $N \geq 20$, the difference between $\pi sd, \nu f_{7/2}p_{3/2}$ (single-shell) and $\pi sd, \nu sdf_{7/2}p_{3/2}$ (multi-shell) results become significant.
This is because the single-shell ground states correspond to excited states in the multi-shell results, which we will revisit later.
In Fig.~\ref{fig:nemgsi_gs} panels (d), (e), and (f), the two-neutron separation energies $S_{2n}$ are shown for neon, magnesium, and silicon isotopes, respectively. Here we see a marked agreement with respect to experiment as the $N=20$ gap is crossed, compared to standard single major shell VS-IMSRG calculations here and in earlier studies~\cite{Simonis2017}.

Turning to excitation energies, the difference between single- and multi-shell results is even more pronounced.
In Fig.~\ref{fig:NeMgSi_Ex}, the first $2^{+}$ energies and the $E2$ transition probabilities from $0^{+}_{1}$ to $2^{+}_{1}$ [$B(E2)$] are shown for the neon, magnesium, and silicon isotopes.
Overall, the calculated $2^{+}$ energies are systematically higher than the experimental data, which has already observed in earlier works~\cite{Simonis2017} and is again likely due to the IMSRG(2) truncation.
The calculated $B(E2)$ values tend to be much smaller than the data, also consistent with earlier findings~\cite{Parzuchowski2017a}.
While ab initio methods based on spherical references tend to underpredict $B(E2)$ values due to missing contributions from many-particle many-hole excitations, the trends are generally well reproduced~\cite{Hend18E2}.

The 2$^{+}$ energies from $sd$ and $\pi sd, \nu f_{7/2}p_{3/2}$ (single-shell) deviate from the experimental trend at $N=20$, while $\pi sd, \nu sdf_{7/2}p_{3/2}$ (multi-shell) calculations show a modest lowering at $N=20$, improving the agreement.
Also, when extending the valence space, $B(E2)$ values increase at $N=20$ for the neon and magnesium isotopes and decrease for silicon.
These improvements at $N=20$ are again because the single-shell results capture the different states in the multi-shell calculations, which will be shown later.
Combining 2$^{+}$ energies and $B(E2)$, we see the clear collapse of the $N=20$ gap in the neon and magnesium and its persistence in silicon.
Overall we find that a proper treatment of cross-shell physics is essential to discuss the disappearance of $N=20$ gap, which is consistent with the discussion with large-scale shell model calculations~\cite{Caurier2014,Tsunoda2017}.
Finally we note that for $N \geq 26$, the $\beta$-dependence of $\pi sd, \nu sdf_{7/2}p_{3/2}$ results becomes sizable, as also seen in Fig.~\ref{fig:nemgsi_gs}.
This suggests that we may need to employ another valence space to discuss the $N=28$ gap, essentially re-centering the fermi surface.

 \begin{figure}[t]
    \centering
    \includegraphics[clip,width=\linewidth]{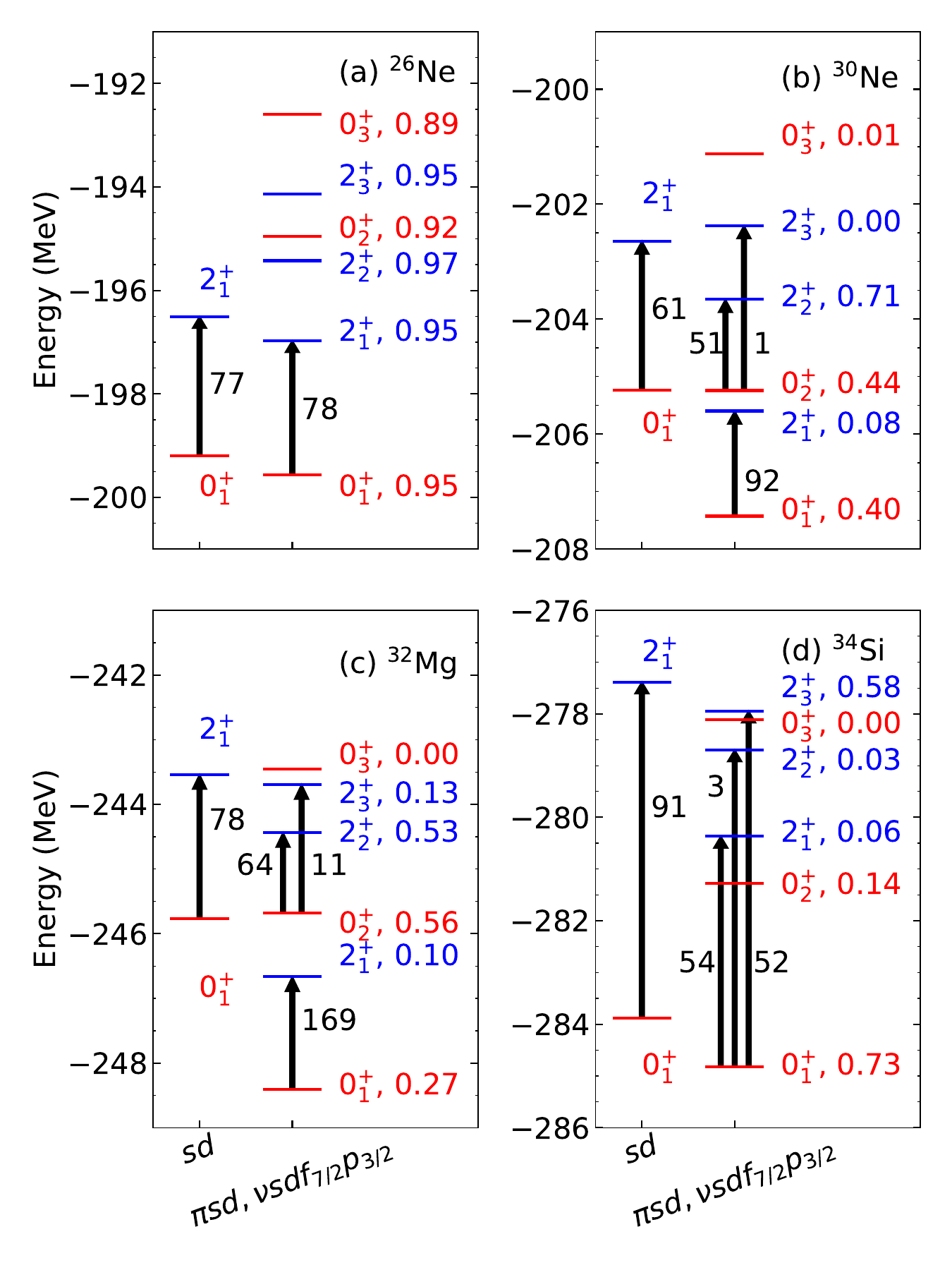}
    \caption{$0^{+}$ and $2^{+}$ states
    for $^{26}$Ne (a), $^{30}$Ne (b), $^{32}$Mg (c), and
    $^{34}$Si (d) obtained from $sd$ and $sdf_{7/2}p_{3/2}$ calculations.
    For $sd$ results, the lowest 0$^{+}$ and $^{+}$ states are shown.
    For $sdf_{7/2}p_{3/2}$ results, the three lowest 0$^{+}$ and 2$^{+}$ states are shown.
    The numbers shown in the $sdf_{7/2}p_{3/2}$ results indicate the
    0$\hbar\omega$ components.
    The numbers nearby arrow indicate $B(E2, 0^{+} \rightarrow 2^{+})$ in
    units of $e^{2}$fm$^{4}$.
    All the calculations are done with $e_{\max}=12$ and
    $E_{3\max}=16$.
    }
    \label{fig:sd_vs_sdpf}
\end{figure}

Beyond $N=20$, the 2$^{+}$ energies from $\pi sd, \nu sdf_{7/2}p_{3/2}$ are still systematically higher than the data, while $\pi sd, \nu f_{7/2}p_{3/2}$ results show very good agreement with data, especially for the silicon isotopes.
The agreement beyond the shell closure was also pointed out in earlier VS-IMSRG work~\cite{Simonis2017}.
We reproduce the $\pi sd, \nu f_{7/2}p_{3/2}$ results when we perform the calculations with $\pi sd, \nu sdf_{7/2}p_{3/2}$ Hamiltonian, with a $0\hbar\omega$ truncation.
To see the difference between the single- and multi-shell calculations more clearly, we show the $0^{+}$ and $2^{+}$ energies and the $B(E2)$ in $^{26}$Ne, $^{30}$Ne, $^{32}$Mg, and $^{34}$Si within the $sd$ and $\pi sd, \nu sdf_{7/2}p_{3/2}$ spaces in Fig.~\ref{fig:sd_vs_sdpf}.
For $^{26}$Ne, outside of the island of inversion, it is clear that the $sd$ results correspond to the lowest two states from the $\pi sd, \nu sdf_{7/2}p_{3/2}$ calculation.
For $^{30}$Ne and $^{32}$Mg, there is not the clear connection as seen in the $^{26}$Ne case.
However, looking at the $0\hbar\omega$ component (i.e. the probability to have zero excitations into the $pf$ shell) and $B(E2)$ values, it seems the $sd$ calculations capture states dominated by the 0$^{+}_{2}$ and 2$^{+}_{2}$ in the $\pi sd, \nu sdf_{7/2}p_{3/2}$ calculations, respectively.
For $^{34}$Si, the $sd$ ground state corresponds to the $\pi sd, \nu sdf_{7/2}p_{3/2}$ ground state.
The $sd$ 2$^{+}_{1}$ seems to capture the strongly mixed states, because the $B(E2)$ values from $sd$ and $\pi sd, \nu sdf_{7/2}p_{3/2}$ are significantly different.
Considering these findings, it appears that the agreement of the $\pi sd, \nu f_{7/2}p_{3/2}$ results with the experimental 2$^{+}$ excitation energies is accidental.
For the intruder-dominant states, the original problems cannot be mapped sufficiently into the single-shell space problems under the IMSRG(2) approximation, and we need to derive the Hamiltonian for the two-major-shell valence space.

\subsection{\label{sec:O16} Excitation spectrum in $^{16}$O}

In the standard VS-IMSRG prescription of choosing single-shell valence spaces, doubly magic nuclei were out of reach except with a single-reference calculation, which could not access excited states. Here we illustrate the utility of multi-shell spaces by calculating the ground-state energy and excited state spectrum of $^{16}$O.
To benchmark with the results from equation of motion coupled-cluster (CC) theory and IMSRG, we employ the N$^{2}$LO$_{\rm sat}$ interaction~\cite{Ekstrom2015}, whose low-energy constants were fitted to reproduce ground-state properties of selected nuclei up to $A=25$.

\begin{figure}[t]
    \centering
    \includegraphics[clip,width=\columnwidth]{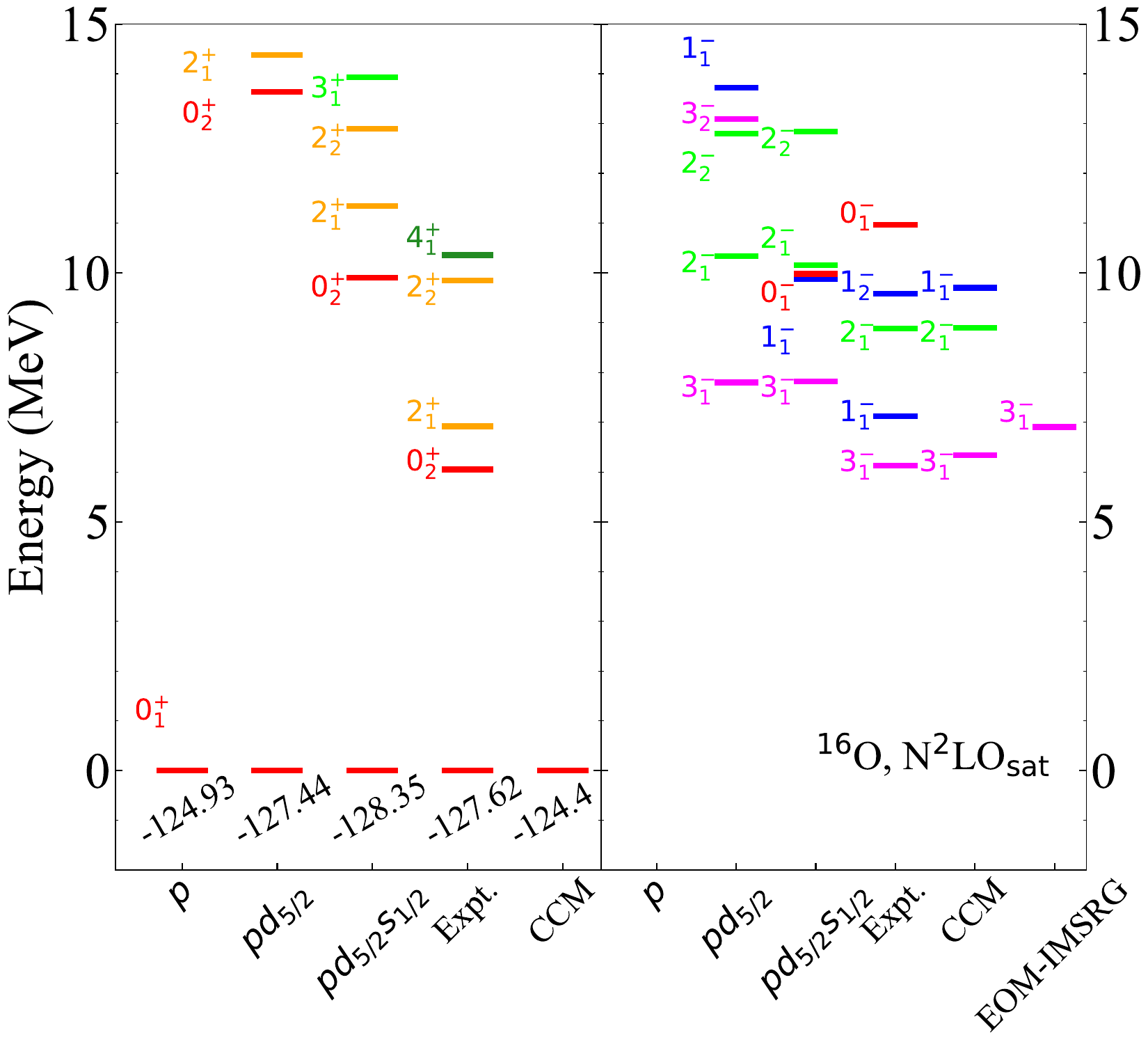}
    \caption{Excitation spectra for $^{16}$O obtained from
    $p$-shell, $pd_{5/2}$-shell, and $pd_{5/2}s_{1/2}$-shell calculations.
    Experimental data are taken from nuclear data~\cite{NationalNuclearDataCenter}.
    The calculation results with coupled-cluster method (CCM) and
    equation-of-motion (EOM) IM-SRG are taken from Refs.~\cite{Ekstrom2015} and
    \cite{Parzuchowski2017a}, respectively.
    The left and right panels show five
    lowest positive and negative parity states, respectively.
    All the calculations are done in the $e_{\max}=12$ model space with
    N$^{2}$LO$_{\rm sat}$ interaction~\cite{Ekstrom2015}}
    \label{fig:O16_spectra}
\end{figure}

In Fig.~\ref{fig:O16_spectra} we show the calculated spectrum for $^{16}$O, taking $^{4}$He as the core and proton-neutron $p$, $pd_{5/2}$, and $pd_{5/2}s_{1/2}$ orbits as the valence spaces.
The calculations are done in a $e_{\max}=12$ and $E_{3\max}=14$ space (where the results are converged) and using $\beta=3$ for the c.m. Hamiltonian term.
Since the $p$-shell orbits are fully occupied, the $p$ valence-space trivially gives the ground state energy.
With an extended valence space, the ground-state energies are seen to decrease by a few MeV.
If the applied IMSRG transformation were perfectly unitary, the two approaches would yield identical results.
This additional lowering is an illustration of the effect of the many-body forces induced by the transformation and neglected in the IMSRG(2) approximation.
The excitation energies of the negative-parity $3^{-}$ and $2^{-}$ states, obtained with $pd_{5/2}$ and $pd_{5/2}s_{1/2}$ valence spaces are very similar.
This indicates that excitations from the $p$-shell orbits to $d_{5/2}$ orbits are dominant, consistent with the earlier CC investigations~\cite{Ekstrom2015}.

The difference in energies from coupled cluster~\cite{Ekstrom2015} and equations-of-motion IMSRG (EOM-IMSRG)~\cite{Parzuchowski2017a} gives an indication the level of the error of the approximations employed in the two methods.
We note that CC with perturbative triples,  CCSD(T), is generally more accurate than the IMSRG(2) framework~\cite{Hergert2016}, and as pointed in Refs.~\cite{Simonis2017,Morris2018,Taniuchi2019}, IMSRG(2) tends to also produce higher first excited-state energies compared to CCSD(T) and experiment.
The comparison of our energies with the EOM-IMSRG indicates the effect of the induced many-body terms by the additional valence-space decoupling, though it could also reflect excitations beyond 2p2h which are missing in the EOM calculation but captured in the diagonalization of the VS-IMSRG Hamiltonian.

Adding the $s_{1/2}$ orbit significantly lowers the energies of higher excited states.
While the importance of $s_{1/2}$ orbits for the $1^{-}$ state was also discussed in the CC calculation, our $1^{-}$ state is slightly lower than the $2^{-}$ state, in contrast to the CC results.
We confirmed that this is indeed caused by three-particle-three-hole (3p3h) contributions which are not captured in Ref.~\cite{Ekstrom2015}; in our calculations, the 1p1h, 3p3h, and 5p5h components are 77\%, 22\%, and 1\%, respectively.
We find that the positive-parity states are dominantly 4p4h.
The 0$^{+}_{2}$ energy linearly rises with increasing $\beta$, while $\langle 0^{+}_{2} | H_{\rm c.m.}| 0^{+}_{2}\rangle$ is nearly zero.
Recalling the Hellmann-Feynman theorem [$dE(0^{+}_{2})/d\beta = \langle 0^{+}_{2} | H_{\rm c.m.}| 0^{+}_{2}\rangle$], this could be another indication that IMSRG(2) does not work to express the state.
Also, these states converge slowly with respect to $e_{\max}$, and are not converged at $e_{\max}=12$.
This is consistent with the findings in Ref.~\cite{Horiuchi2014} that nucleon excitations much higher than those taken in this work would be needed to express the $0^{+}_{2}$ state.

 \begin{figure}[t]
    \centering
    \includegraphics[clip,width=\columnwidth]{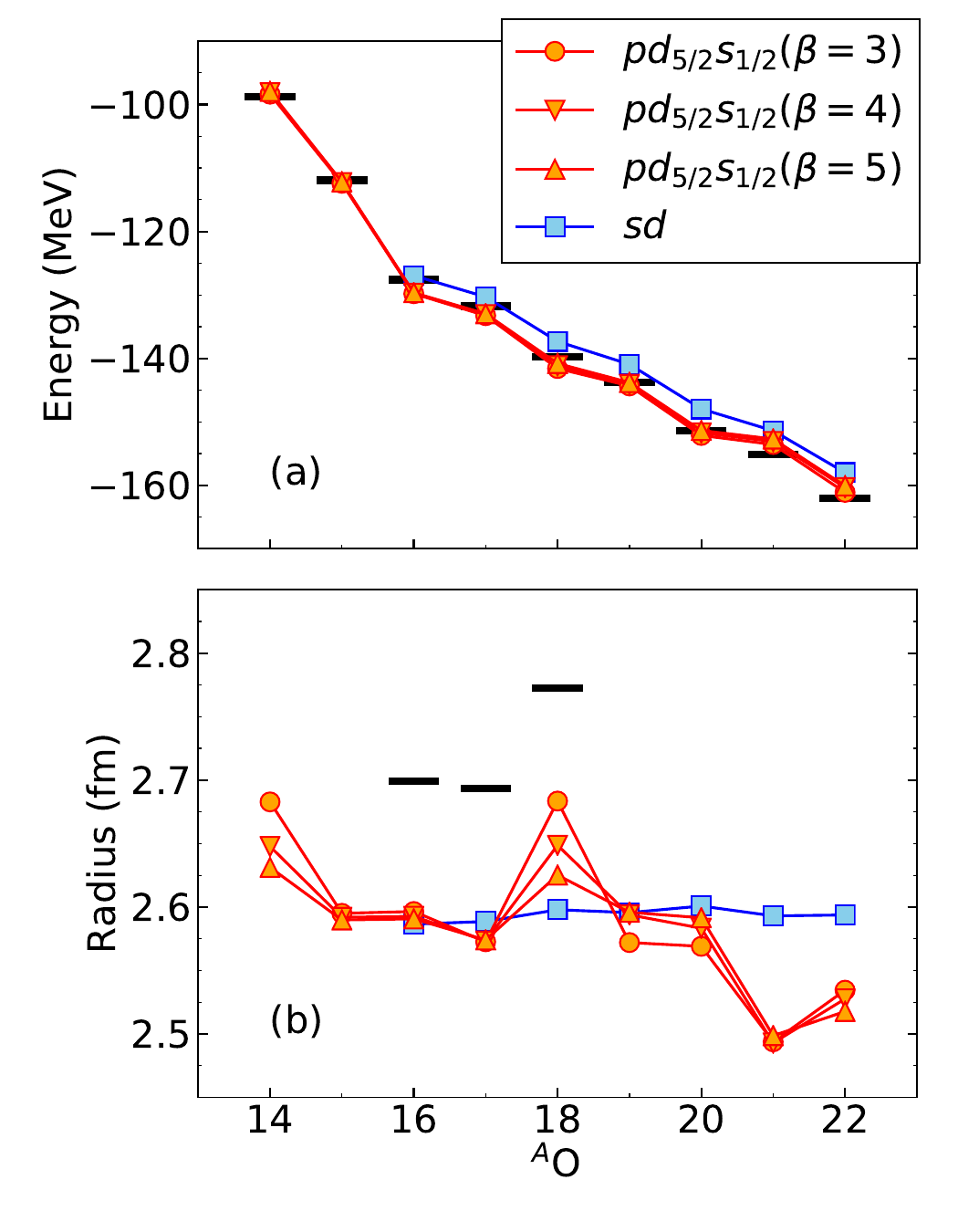}
    \caption{Ground-state energies (a) and charge radii (b) for oxygen isotopes.
    The experimental data (black bars) are taken from Refs.\cite{Wang2017,Angeli2013}.
    All the calculations are done with $e_{\max}=12$ and $E_{3\max}=16$.}
    \label{fig:O_gs}
\end{figure}

\subsection{\label{sec:oxygens} Oxygen and calcium isotopes}

As in many ab initio studies~\cite{Hagen2012a,Cipollone2013,Hergert2013,Cipollone2015}, the oxygen isotopic chain are also an ideal playground to examine multi-shell effective Hamiltonians.
Here, we use $^{16}$O-core $sd$ and $^{4}$He-core $pd_{5/2}s_{1/2}$ valence spaces with the 1.8/2.0 (EM) interaction discussed above for $e_{\max}=12$ and $E_{3\max}=16$.
In Fig.~\ref{fig:O_gs} (a), the ground-state energies of oxygen isotopes are shown.
First we notice that the $\beta$-dependence of the ground-state energies are negligibly small.
Additionally, decoupling of the $p$ orbits gives more binding throughout the chain, due to the IMSRG(2) truncation error.
On the other hand, the $\beta$-dependence of the charge radii, shown in panel (b) of Fig.~\ref{fig:O_gs}, is not as clean.
Except for $^{14,18}$O, the converged radii can be found by increasing $\beta$.
Since the many-body induced terms would be more serious for larger $\beta$, we stop at $\beta=5$.
Although our $^{18}$O charge radius is still $\beta$-dependent, the experimental trend $R_{\rm ch}(^{18}{\rm O}) > R_{\rm ch}(^{16}{\rm O}) > R_{\rm ch}(^{17}{\rm O})$ is likely coming from the excitations of nucleons from $p$ to $sd$ orbits, as expected.
The dip for $A$$>$20 is qualitatively consistent with recent experimental results from proton-scattering data~\cite{Kaur2018}.
Since the sensitivity to $\beta$ depends on the nuclide and operator, we need to investigate further to obtain reliable results.

  \begin{figure}[t]
    \centering
    \includegraphics[clip,width=\columnwidth]{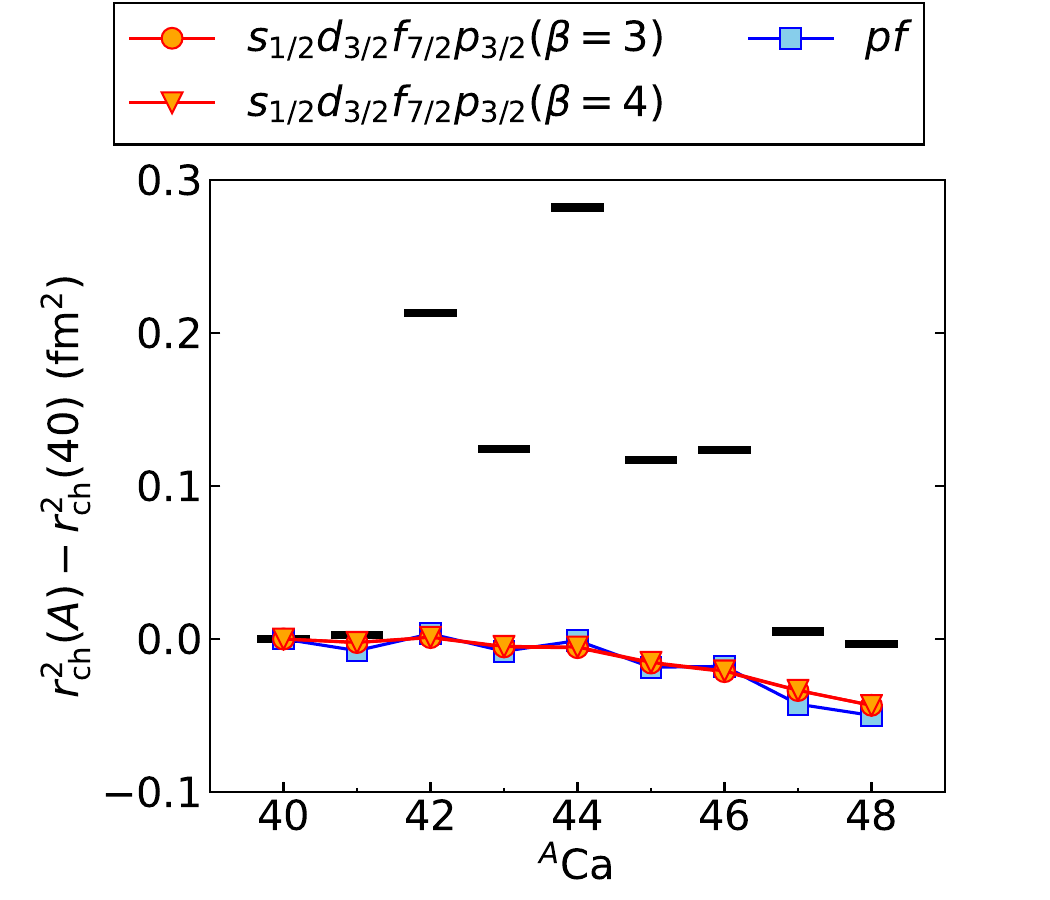}
    \caption{Isotope shift of charge radii for calcium isotopes.
    The experimental data (black bars) are taken from Ref.~\cite{Angeli2013}.
   All the calculations are done with $e_{\max}=12$ and $E_{3\max}=16$.}
    \label{fig:Ca_gs}
\end{figure}

The charge radii of calcium isotopes are also worth investigating, not only because of the recent exploration to the neutron rich region~\cite{GarciaRuiz2016}, but it is also an outstanding challenge for theory to reproduce the parabolic isotope shift trend from $^{40}$Ca to $^{48}$Ca, particularly with ab initio methods.
In earlier shell-model calculations with an $s_{1/2}d_{3/2}f_{7/2}p_{3/2}$ valence space above a $^{28}$Si core~\cite{Caurier2001}, it was claimed that the isotope shift and other observables provided evidence that the $Z=20$ shell closure is incomplete in the calcium isotopes.
We show the isotope shift of the charge radii for calcium isotopes using the $^{28}$Si-core $s_{1/2}d_{3/2}f_{7/2}p_{3/2}$ spaces in Fig.~\ref{fig:Ca_gs}.
Note that the absolute charge radii are $\beta$-independent and consistently smaller than the experimental data.
As we move to $^{48}$Ca, we observed that the nucleon excitations from the $sd$ to $pf$ ($n_{sd \rightarrow pf}$) are suppressed;
for example $n_{sd \rightarrow pf}(^{40}{\rm Ca})=1.29$,
 $n_{sd \rightarrow pf}(^{44}{\rm Ca})=1.08$, and
 $n_{sd \rightarrow pf}(^{48}{\rm Ca})=0.39$ at $\beta=3$.
Although our calculation shows non-negligible excitations from $sd$ to $pf$ in the lighter isotopes, the shell closure is restored when the neutrons are added, in contrast\footnote{Again, occupation numbers are not observable and so calculations employing different Hamiltonians need not agree on them.} to the earlier shell-model calculations~\cite{Caurier2001}.
In Figure~\ref{fig:Ca_gs},
 the isotope shifts from our calculations are flat as a function of $A$ in sharp contrast with the experimental trend.
Also, the agreement of the $pf$ and $s_{1/2}d_{3/2}f_{7/2}p_{3/2}$ results indicate that
 the effects of the excitations from $sd$ to $pf$ on radii are somehow canceled out.
To investigate further, it would be helpful to compare with other ab initio calculation methods, as this result is indeed unexpected.

\section{\label{sec:summary} Conclusion and outlook}
In this paper we used the VS-IMSRG to derive the first multi-shell valence-space Hamiltonians from ab initio theory.
To make this feasible we added an energy shift $\Delta$ in the denominator of the generator $\eta$ of IMSRG flow equation to avoid the single-particle level crossing that otherwise occurs during the evolution.
Although we can now in principle derive effective Hamiltonians for general multi-shell valence spaces, in some cases adding the c.m. Hamiltonian spoils the hierarchy of the IMSRG induced many-body terms.
In practice, therefore, we have to choose carefully the valence space so that this hierarchy is preserved.

We then used these multi-shell valence-space Hamiltonians to provide an ab initio description of the island of inversion in the region above oxygen. Here we generally reproduce the expected evolution of the $N=20$ magic number in the neon, magnesium, and silicon isotopes and see the prominence of intruder configurations and cross-shell excitations in ground and excited states.
We then benchmark our calculations against other large-space ab initio methods in the excitation spectra of $^{16}$O, showing reasonable agreement with the earlier calculations starting with the same NN+3N  interaction.
Finally, we explored the ground-state energies and radii of oxygen and radii isotope shift of calcium isotopes, finding that in order to obtain reliable results for a given nuclide and a given observable, we need to check the center of mass contamination by varying the parameter $\beta$.

This work further extends the reach of the VS-IMSRG to regions near major oscillator shell closures, where artifacts were clearly evident in previous calculations.
Since in this framework we can freely add or remove the valence orbits, we can investigate what are the essential degrees of freedom to describe a many-body wave function (under the IMSRG(2) approximation).
While diagonalization of multi-shell Hamiltonians often becomes computationally challenging towards the $sd-pf$-shell, this method nevertheless opens the way for ab initio investigations to improve the trend of calcium charge radii, explore the island of inversion above calcium, generate multi-shell Hamiltonians useful for neutrinoless double beta-decay calculations~\cite{Jiao170v}, and guide future experimental efforts exploring neutron rich nuclei with rare isotope beams.

\begin{acknowledgments}
We would like to thank B.~R. Barrett, J. Men\'endez and Y. Suzuki for enlightening discussions.
We are grateful to J. Simonis and P. Navr\'atil for providing the 1.8/2.0 (EM) and N$^{2}$LO$_{\rm sat}$ matrix element files.
TRIUMF receives funding via a contribution through the National Research Council of Canada. This  work was  supported by NSERC, the Arthur B. McDonald Canadian Astroparticle Physics Research Institute, and the US Department of Energy (DOE) under contract DE-FG02-97ER41014.
NS acknowledges
Priority issue 9 to be tackled by using Post K Computer from MEXT and JICFuS, Japan. Computations were performed with
an allocation of computing resources on Cedar at WestGrid and Compute Canada, and on the Oak Cluster at TRIUMF managed by the University of British Columbia department of Advanced Research Computing (ARC).

\end{acknowledgments}



%

\end{document}